\documentclass[
rmp,
fleqn,nofootinbib,reprint]{revtex4-2}
\makeatletter \renewcommand\Dated@name{} \makeatother % Date-Format Reformation
\usepackage{Suppltorevtex} %A package of personal preambles

\newcommand*{\blue}[1]{{\color{Blue}#1}}
\newcommand*{\red}[1]{{\color{Red}#1}}

\newcommand*{\Ospin}{{\protect \substack{\text{{\footnotesize \faCircle}}}}}
\newcommand*{\Pspin}{{\protect\substack{\text{\footnotesize\color{Red}\faPlusCircle}}}}
\newcommand*{\Mspin}{{\protect\substack{\text{\footnotesize\color{Blue}\faMinusCircle}}}}

\begin{document}
% \rmfamily
\title{Rod magnets inscribed in an elastic cuboid\\
\large \textit{Interpreting single-domain ferroics in Onsager's spirit}}
\author{\href{mailto:jaiman.ng@gmail.com}{Jimin WU}}
\affiliation{
South Bay Interdisciplinary Science Centre, Songshan Lake Materials Laboratory,
\\523830 Dongguan, China\looseness=-2
}
\begin{abstract}
A toy model for ferroic orders through entropy. As the rod/spin concentration (or the volume enclosing the ferroic rods) is variable by a tunable force conjugate to the order parameter, the model exhibits hysteresis associated to some discontinuous isotropic-nematic transitions explained essentially by \textsc{Onsager}'s hard-rod model. The uniaxial anisotropy and nonperiodicity by construction is reminiscent of \textsc{Stoner–Wohlfarth}'s single-domain magnet, and the toy model may ultimately be seen as a minimalistic for hard ferroics with some possible self-averaging disorders. Spin togglings shall be reliable due to the clean two-well energy landscapes in the athermal transition theory.
\end{abstract}

\maketitle
% \tableofcontents

\section{Quantum fluctuations\\
\footnotesize \textit{The philosophy in athermal transitions}}

\textsc{Onsager}'s model \cite{onsager1949effects} of athermal transition through rival entropies consists of rigid rods (referred to as \textsc{Onsager}'s rods hereafter) of infinite length-diametre ratio $\epsilon^{-1} \coloneqq l/d$. The system shall be \textit{isotropic} (with $[1+s]$ different axes along which the rods tend to align) and \textit{nematic} (all rods align uniaxially) respectively at two extremum concentrations \begin{math}
  c \coloneqq C/V
  % \sim \mlr{\epsilon^{\pm 1-1}}{l^3}
\end{math},\footnote{
I shall stick to the notational conventions in apps.~\ref{app:BLtheo} and \ref{app:statprefmaxientr&minifreeener}, in which some elements of statistical physics are `popularised' for interested audiences.
} or equivalently, packing fractions \cite{parisi2020theory} \begin{math}
  \varphi \coloneqq 2dl^2 c \sim
  % \mlr{\qty(5\pm 1)\epsilon^{\pm 1}}{4\pi}
  \epsilon^{\pm 1}
\end{math}.\footnote{
Imagine two perpendicular rods excluding each other's geometrical centres from occupying a certain volume, the average \emph{excluded volume} produced by a single rod $\sim 2dl^2$; \textit{cf.}\ fig.~1.a of \cite[\textbf{b}]{frenkel2000&2015}. Other examples of excluded volumes could be
(i) \begin{math}
  \mlr{4 \pi l^3}{3}
\end{math} for a radius-$l$ sphere
(ii) \begin{math}
  2 \pi d^2 l
\end{math} for a cylinder of length $2l$ and diametre $2d$, with packing fractions \begin{math}
  \sim \mlr{\qty(5\pm 1)\epsilon^{\pm 1}}{4\pi} 
\end{math} respectively.
}
To locate the transition point $\varphi^*$ between the least and the most concentrated phases, one may refer to the energy/entropy difference\footnote{
\textit{Cf.}\ app.\!~\ref{app:statprefmaxientr&minifreeener}; app.\!~\ref{app:misccalc} shall elaborate somewhat the sketch of the model here.
} \cite{frenkel2000&2015}
\begin{eqnarray}
  &&\ub{\textstyle
    \ln\qty{
      V - 2 \pi d^2l C
    } + \ln{1}
  }_{
    S_{\varphi_{\text{nem}}}
    = S_{\varphi_{\text{nem}}}^{\text{translational}} + S_{\varphi_{\text{nem}}}^{\text{rotational}}
  } - \ub{\textstyle\bigl(
    \ln\qty{
      V - 2 dl^2 C
    } + \ln\qty{1+s}
  \bigr)}_{
    S_{\varphi_{\text{iso}}}
    = S_{\varphi_{\text{iso}}}^{\text{tra}} + S_{\varphi_{\text{iso}}}^{\text{rot}}
  } \eqqcolon \notag\\&&\delta{S}_\varphi = \ln{\frac{
    1 - \epsilon \pi \varphi
  }{
    \qty(1 - \varphi)(1+s)
  }} \xRightarrow[
    \epsilon \rightarrow 0
  ]{
    0 \equiv \delta{S}_{\varphi = \varphi^*}
  } \varphi^* = \frac{s}{1+s} \begin{cases}
    \xrightarrow{s\rightarrow 0} 0,
    \\\xle{s = 1} 1/2,
    \\\xrightarrow{s\rightarrow \infty} 1,
  \end{cases}
\label{eq:del{S}}\end{eqnarray}
where
(i) \begin{math}\qty(
  V - 2\pi d^2 l C
)\end{math} is the explorable space for rods in a nematic single phase, and \begin{math}\qty(
  V - 2 d l^2 C
)\end{math} the one in an isotropic case
(ii) $(1+s)$ is interpreted as number of component species of a self-assembly in \cite[\textbf{b}]{frenkel2000&2015}
(iii) we shall call a $\varphi \sim \varphi^*$ \textit{semidilute} \cite{doi1988theory}, and the transition depends trivially on the exact value of $\qty[\mlr{s}{(1+s)}]$ if it remains some finite constant.

Summaries of the essence and pedagogical practices have been abundant \cite{frenkel2000&2015}--\cite{vroege1992phase} since \textsc{Onsager}'s first insight. Three basic facts about the model are
\begin{description}[wide=0pt]
\item[Athermality]
as \textsc{Mayer}'s function used to polynomial/\textsc{Virial} expand a partition function depends trivially on temperatures (footnote~\ref{foot:Mayefct}); this shall be rephrased in the language of quantum fluctuations in a moment
\item[Discontinuous phase transition]
whose many links to the ferroic hystereses\footnote{
  Below some critical/\textit{order disruption} temperatures of \textsc{Curie}'s, \textsc{N\'eel}'s \textit{etc.} \cite{aharoni2000introduction}.
}
are in evidence
\item[Two-well potential]
as higher-than-second-order \textsc{Virial}'s terms are greatly suppressed by $\epsilon$, the free energy $f[\varphi]$ shall have a two-well landscape by \textsc{Landau}'s mean-field (m.f.) idea in the close vicinity of the semidilute transition point; if the isotropic minima\footnote{
Strictly speaking, of some \begin{math}
  \delta{f}_{\varphi}\qty[\text{order parameter }\sigma]
\end{math} now for different m.f.\ solutions $\varphi$; but the m.f.\ constraint/saddle-point equation sets some relations between $\sigma$ and its conjugate force $\varphi$, so everything could eventually be expressed as a function of $\varphi$ (or $\sigma$). \textit{Cf.}\ app.\!~\ref{app:misccalc}.
}
are pinned, then any \textit{errors} in a desired micro-state pattern can only concern the nematic ones.
\end{description}
The essay's goal is to encapsulate ultimately \textsc{Onsager}'s spirit into ferroics, with the trial prototype here \textsc{Stoner-Wohlfarth}'s (S-W's) macro-spin \cite{tannous2008stoner,carrey2011simple}. \textit{Viz.}, to fashion a ferroic switch triggered by a variable field with the usual outcome of polarisation reversals, but the internal mechanism governed by \textsc{Onsager}'s entropy competitions---hence athermal, and a reliable two-state toggling free from pitfalls of some muddled energy landscapes.

Before continuing the main thread of narration, I shall give a few more legitimacy checks on the aforementioned model features and some philosophies involved.
\begin{description}[wide=0pt]
\item[Quantum phase transition]{
stands out typically when phasing out the `internal energy-entropy' tradeoff in eq~\eqref{def:S} at zero temperature, with frozen particles and proliferating quantum fluctuations by \textsc{Heisenberg}'s uncertainty principle.\footnote{
Some \textsc{Schwarz}'s/triangle inequality, \cite{weinberg2015quantum} such that a vanishing root-mean-square deviation $\Delta_{\Psi} \vbx$ of the position operator $\vbx$ evaluated at a given system state $\Psi$
\begin{math}
\xRightarrow[\text{uncertainty relation}]{
  \Delta_{\Psi} \vbx . \Delta_{\Psi} \vbp \geq \text{constant}
}\end{math} momentum root-mean-square \begin{math}
  \Delta_{\Psi} \vbp \rightarrow \infty
\end{math}.
} In \textsc{Onsager}'s model, the rod exploration space or positional uncertainty is abruptly reduced when $\varphi$ drops across $\varphi^*$, leading to multiplication of the quantised rotational degrees of freedom,\footnote{
Sending centres of all rods to a fixed $\bbR^3$-origin, $\bbR^3$-rotations of the directional unit vectors $\vbn$ are then typified by group elements $\msR \in$ SO$_3$. Note some $\bbC^2$-`spin rotations' in quantum mechanics can be represented by
\begin{math}
  \msU_{\theta} = \me^{\mlr{
    \theta \vbn . \vv{\sigma}
  }{2\mi}}
  \in 
\end{math} SU$_2$ generated by \textsc{Pauli}'s matrices \begin{math}
  \sigma_{x,y,z}
\end{math}, such that SU$_2$($\simeq$ a unit $3$-sphere S$^3$, topologically) acts locally like SO$_3$($\simeq$ quotient group \begin{math}
  \mlr{{\text{S}}^3}{\qty{\pm 1}}
\end{math}).
}
and thus some athermal `quantum phase transition' like those of quantum \textsc{Ising}/rotor models \textit{etc.}\ in \cite{sachdev2006&2011}.
}\item[Mean-field theory\label{sec:meanfieltheo}]{\!\!\footnote{
\textit{Cf.}\ chaps. 1--2 of \cite{cardy1996scaling}, and \cite{parisi2020theory} \textit{etc.}
}
(i) coarse-grains fractions of a statistical system into self-equilibrial `molecules'
\begin{equation*}
  m^i \coloneqq \expval{\textstyle
    \sigma^i = \delta{\sigma^i} + \qty(
      m \equiv \sum_{j=1}^C \mlr{m^j}{C}
    ) 
  }_{\mcZ},
\end{equation*}
with $\sigma^i$ a variable (named \textit{spin}) representative of fraction $i$
(ii) approximates the partition functions/statistic $\mcZ$ by a series of the local means. Such m.f.\ predictions may break down by \textsc{Ginzburg}'s criterion \cite{cardy1996scaling} for a low-dimensional system, with finite-range interactions, at a \textit{continuous}\footnote{
In 1st derivatives of the free energy with respect to applied forces.
}
phase transition where the correlation length $\xi$ (characteristic size of fluctuations \begin{math}
  \expval{\delta{\sigma^i}.\delta{\sigma^j}}
\end{math}) explodes.\footnote{
\textit{I.e.}, nonnegligible fluctuations over all distances force the entire system to form a unique critical phase; the properly compatible ideas of scaling and renormalisation are then needed for quantitative descriptions.
}
As for a \emph{discontinuous} transition like \textsc{Onsager}'s athermal one, an instability of the continuation from one coexisting phase to another shall yield some \textit{hysteresis}/\textit{memory effect}. \textit{I.e.}, the system needs time to readjust between single phases, with $\xi$ finite at such transitions in general. Note the expected transitions in the toy model to be proposed are \textit{abrupt}. As we are not concerned anyway with the asymptotic exponents of critical behaviours during the abrupt periods, a first qualitatively m.f.\ treatment seems adequate.

One more thing to stress is that a uniform \begin{math}
  m \equiv m^{\forall i}
\end{math} is \textit{not} a necessity for optimising a m.f.\ free energy. The implicit homogeneity from a m.f.\ idea is explicitly lost in a disordered system. \textsc{Parisi}'s remedy \cite{parisi2020theory} is to
(i) \textit{replicate} the statistics $\mcZ$ $s$ times and allow `$\sigma^i_a$-$\sigma^j_b$' interactions, but every ($1+s$) replicas $\sigma_{a = 1,2,\dots,1+s}^{\forall i \in \qty{1,\dots,C}}$ form a `molecule' bound by a strong-localisation constraint,\footnote{
Reminiscent of the $s$ extra rotational degrees of freedom for each \textsc{Onsager}'s rod in an isotropic phase than in a nematic one.
} with $
  \qty{\sigma_{a=1}^i
  \equiv \sigma^i}_{i=1}^C
$ the reference \textit{quenched} disorder; such molecular liquid shall recover some translational and rotational invariance, and one could use again methods like \textsc{Virial}'s expansion \textit{etc.}\ generalised from theories of conventional atomic/molecular liquids
(ii) \textit{conflate} dissenting statistics into a full average/integral $
  \mcZ^{1+s} \sim \prod_{a=1}^{1+s} \mcZ_a
$, and the final desired \textit{glass free energy} is an analytic continuation of \begin{equation*}
  - \mlr{\qty(
      \partial_s \mcZ^{1+s}
    )}{\mcZ C}
  \xrightarrow[]{s\rightarrow 0} \mlr{\qty(
    - \ln{\mcZ}
  )}{C}
  = f_{\text{gla}}.
\end{equation*}
}
\end{description}

\section{Phenomenologies for hystereses}
\subsection{A thought experiment}
The task is simply to well encapsulate $C$ magnetised \textsc{Onsager}'s rods in an elastic cuboid\footnote{
Or any similar geometry like a cylinder of length $L$ and diametre $D$, for negligible boundary effects.
}
of height $L \sim l$ and volume $D^2 L \sim l^3$. Small variations of the cuboid volume/occupied volume of the rods allow variations of packing fractions \begin{math}
  \varphi
\end{math} around the critical \begin{math}
  \varphi^*
\end{math} ($\equiv \mlr{1}{2}$ for simplicity) in eq~\eqref{eq:del{S}}.\footnote{
The average nematic excluded volume for rods in the cramped box is now a cylinder $\pi d^2 l$ of length $l$ (rather than $2l$) and diametre $2d$, but this will not change the final \begin{math}
  \eval{\varphi^* \textstyle}_{\epsilon \rightarrow 0}
\end{math} anyway.
}
Label the three axes through the centres of the opposite cuboid faces as $x$, $y$ and $z$; let the vertical axis be $z$, so that the principal fa\c{c}ade and the aerial view are respectively in the $x$-$z$ and $x$-$y$ planes.
\subsubsection{\label{sec:initconf}
Initial configuration
\\\textit{Introduction of a mechanism for protecting rods' impenetrability}}
\begin{enumerate}[wide=0pt,label=(\roman*)]
\item all north poles $\Pspin$ are perfectly up in a strong applied force $\vbB \equiv \vbB^z$ (table~\ref{tab:iden}), inducing naturally a unique \textit{nematic axis} as the $z$-axis
\item the horizontal `$\Pspin\hspace{-3pt}\leftrightarrow\hspace{-3pt}\Pspin$' and `$\Mspin\hspace{-3pt}\leftrightarrow\hspace{-3pt}\Mspin$' repulsions keep the confined rods from collapsing together but aligning at equal spacings \begin{math}
  a \lesssim 2 \sqrt{\epsilon} l \ll D
\end{math}, as the preset \begin{itemize*}[label=\textcolor{gray}{$\diamond$}]
\item \begin{math}
  L \equiv l
\end{math}
\item $\qty[\mlr{D}{l}]$ finite
\item \begin{math}
  \varphi = \qty[\mlr{2dl^2\qty(\mlr{D}{a})^2}{D^2 L}]
  \gtrsim \varphi^*
\end{math} for now; note that \begin{math}
  a \gg d = \epsilon l
\end{math}, `\begin{math}
  \uspin \!\!\cdots\!\! \uspin
\end{math}' is hence morphologically nothing more than a bundle of fibres/lines of induction
\end{itemize*}
\item dipolar interactions (\begin{math}
  \sim \mlr{1}{\qty|\vbr^{i,j}|^3}
\end{math} for parallel dipoles $i$ and $j$ at a distance $\vbr^{i,j}$)\footnote{
\textit{Cf.}\ prob.\!~6.21 in \cite{griffiths2017introduction} for a general potential between angled dipoles.
} are assumed to be so weak and well \textit{screened} (\cite{altland2010condensed}, say the characteristic length of damping is \begin{math}
  \sim a
\end{math}) that they compare minorly with the prescribed two-well potential in the system free energy $\mcF$; \textit{cf.}\ discussions of the non-local forces \textit{etc.}\ in some \textsc{Langevin}'s equation \begin{math}
  \partial_t \phi = \qty(- \fdv*{\mcF}{\phi})
\end{math} of \cite{jaglajagla2005hysteresis}, with $\phi(\vbx,t)$ a scalar variable forming the phenomenological $\mcF[\phi]$.
\end{enumerate}
\begin{table}[b]
\caption{\label{tab:iden}
Ferroic dipoles $\vbM$ subject to their conjugate force $\vec{B}$. The varying $\vbB$ is along a fixed vertical axis, on which the projection of the sum of dipoles is represented by a macro-spin
(\textcircled{1}--\textcircled{3}) $\Uspin$\;$\simeq$ \begin{math}
  \uspin \!\!\cdots\!\! \uspin
\end{math}
(\textcircled{5}--\textcircled{7}) $\Dspin$\;\begin{math}
  \simeq \dspin \!\!\cdots\!\! \dspin
\end{math}
(\textcircled{4}, \textcircled{8}) ${\protect \substack{\text{$\Ospin$}}}$\;\begin{math}
  \simeq \mlr{\qty(
    \Uspin + \Dspin
  )}{2}
\end{math}.
}
\begin{ruledtabular}
\begin{tabular}{lcccccccc}
&\textcircled{1}&\textcircled{2}&\textcircled{3}&\textcircled{4}&\textcircled{5}&\textcircled{6}&\textcircled{7}&\textcircled{8}\\\colrule
&\blue{-}&0&\red{+}&\red{+}&\red{+}&0&\blue{-}&\blue{-}\\\bottomrule
$\vbM^{\text{total}}$
& \multicolumn{3}{|@{}c@{}|}{
  $\Uspin$
}&${\protect \substack{\text{$\Ospin$}}}$&
\multicolumn{3}{|@{}c@{}|}{
  $\Dspin$
}&${\protect \substack{\text{$\Ospin$}}}$\\\toprule
$\vbB$ & $\Uparrow$ & 0 & $\Downarrow$ & $\Downarrow$ & $\Downarrow$ & 0 & $\Uparrow$ & $\Uparrow$\\
&\red{+}&0&\blue{-}&\blue{-}&\blue{-}&0&\red{+}&\red{+}
\end{tabular}
\end{ruledtabular}
\end{table}
\begin{figure}[htbp]
\begin{tikzpicture}[baseline={(current bounding box.center)}]
\begin{feynman}[small]
\vertex (o){0};
\vertex[left = 1.5cm of o](x1);\vertex[right = 1.5cm of o](x2){$\vbb$};\vertex[below = 1.5cm of o](y1);\vertex[above = 1.5cm of o](y2){$\vbm$};
\vertex[left = .5cm of o](x3){$\Downarrow$};\vertex[right = .5cm of o](x4){$\Uparrow$};
\vertex[left = 1cm of o](d){\textcircled{4}};\vertex[right = 1cm of o](h){\textcircled{8}};\vertex[below = 1cm of o](f){\textcircled{6}};\vertex[above = 1cm of o](b){\textcircled{2}};
\vertex[right = 1cm of b](a){\textcircled{1}};\vertex[left = 1cm of b](c){\textcircled{3}};
\vertex[left = 1cm of f](e){\textcircled{5}};\vertex[right = 1cm of f](g){\textcircled{7}};
\diagram*{
(x1)--[scalar, gray](x2),(y1)--[scalar, gray](y2),
(a)--[fermion](b)
  --[fermion](c)--[fermion](d)--[fermion](e)--[fermion](e)--[fermion](f)--[fermion](g)
  --[fermion](h)--[fermion](a),
};\end{feynman}\end{tikzpicture}
\caption{
Sketch of key points in a ferroic hysteresis loop of the thought experiment depicted in table~\ref{tab:iden}. We prefer in physics dimensionless variables \begin{math}
  \vbm \coloneq \mlr{\vbM^{\text{tot}}}{\vbM_{\text{sat}}^{\text{tot}}}
\end{math} and \begin{math}
  \vbb \coloneqq \mlr{\vbB}{\vbB_{\text{coe}}}
\end{math}, with $\vbM_{\text{sat}}^{\text{tot}}$ and $\qty(-\vbB_{\text{coe}})$ respectively the \textit{saturation} magnetisation and \textit{coercive} force at \textcircled{1} and \textcircled{4}. The (major) loop is a perfect square for S-W's macro-spin \cite{tannous2008stoner,carrey2011simple} in a varying $\vbB$ along the anisotropy axis, \textit{i.e.}, \begin{math}
  1 \equiv \vbm_{\textit{remanent}}
\end{math} at \textcircled{2}, and \begin{math}
  1 \equiv \vbm_{\text{transition}}
  = - \vbb_{\text{tra}}
\end{math} at \textcircled{3}.
\label{fig:hysteresis}}\end{figure}
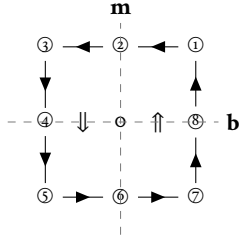
A small translational deviation of a single rod in the above equi-spacing configuration shall raise $\mcF$ by
\begin{eqnarray*}
  &&\delta \mcF \propto \sum_{s= \pm} \qty{
    \sqrt{\qty[
      (x-sa)^2 + y^2
    ]^{-3}} + \sqrt{\text{`$x \leftrightarrow y$'}}
  } - 4a^{-3}
  \quad\notag\\&&\geq 0\qc
  x^2+y^2 \leq a^2 < \qty(|x| - a)^2 + \qty(|y| - a)^2
  ,
\end{eqnarray*}
and makes the system less stable. This malposition amounts to strengthening the effective couplings of adjacent (di)poles in fig.\!~\ref{fig:aeriview} (or shrinking the whole lattice a bit, keeping all dipoles parallel) on a coarse-grained level. Introducing such `pan-tilt' randomness/disorders seems interesting \cite{wiese2021theory} but not the exact task of fashioning for now a simplest m.f.\ toy model.
\begin{figure}[htbp]
\centering
\begin{equation*}
\begin{tikzpicture}[baseline={(current bounding box.center)}]
\begin{feynman}[small]
\vertex (o){$\Pspin$};
\vertex[left = 1cm of o](a1);\vertex[right = 1cm of o](a2){$x$};\vertex[below = 1cm of o](b1);\vertex[above = 1cm of o](b2){$y$};
\vertex[left = .7cm of o](x1){$\Pspin$};\vertex[right = .7cm of o](x2){$\Pspin$};\vertex[below = .7cm of o](y1){$\Pspin$};\vertex[above = .7cm of o](y2){$\Pspin$};
\diagram*{
(a1)--[scalar, gray](a2),(b1)--[scalar, gray](b2),
};\end{feynman}\end{tikzpicture}
\longrightarrow \begin{tikzpicture}[baseline={(current bounding box.center)}]
\begin{feynman}[small]
\vertex (o){};
\vertex[left = 1cm of o](a1);\vertex[right = 1cm of o](a2){$x$};\vertex[below = 1cm of o](b1);\vertex[above = 1cm of o](b2){$y$};
\vertex[left = .7cm of o](x1){$\Pspin$};\vertex[right = .7cm of o](x2){$\Pspin$};\vertex[below = .7cm of o](y1){$\Pspin$};\vertex[above = .7cm of o](y2){$\Pspin$};
\vertex[right = .2cm of o](x3);\vertex[above = -.05
cm of x3](z){$\Pspin$};
\diagram*{
(a1)--[scalar, gray](a2),(b1)--[scalar, gray](b2),
};\end{feynman}\end{tikzpicture}
\end{equation*}
\caption{
Aerial view of a deviation of the central pole.
\label{fig:aeriview}}\end{figure}
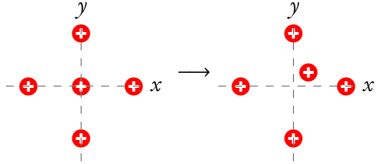

\subsubsection{Interpreting the hysteresis in Onsager's spirit
\\\textit{Forging variations of concentration by tuning the force applied}}
Assuming that the above protection for impenetrability remains effective,

$\vbb_{\textcircled{1}}$$\rightarrow$$\vbb_{\textcircled{2}}$\textbf{:}
\textsc{Zeeman}'s energy, \begin{math}
  - \vbb. \vbm \xle[
    \qty|\vbb| \eqqcolon b \equiv b_z
  ]{
    \qty|\vbm| \eqqcolon m \equiv m_z
  } - bm
\end{math}, is reduced from $1$ to $0$. \textit{Viz.}, the system need less local anisotropy energy (\textit{i.e.}, the prescribed two-well potential) or dipolar interaction to compensate the \textsc{Zeeman} term. Following the same line of reasoning about fig.~\!\ref{fig:aeriview}, the system shall expand (keeping all dipoles parallel), forging a reduction in $\varphi$, while $1 \equiv m$ as the number of rods and the strength \begin{math}
  \qty|\vbM \equiv \vbM^z|
\end{math} of a single dipole both remain identical by construction.

$\vbb_{\textcircled{2}}$$\rightarrow$$\vbb_{\textcircled{3}}$$\xrightarrow{\vbb_{\textcircled{4}}}$$\vbb_{\textcircled{5}}$\textbf{:}
the same reduction continues untill a breakdown at \textcircled{3}, where $\varphi$ touches the transition $\varphi^*$, and all rods flip in \textsc{Onsager}'s spirit at once to the configuration \textcircled{5} with $\vbm$'s direction the same as $\vbb$'s again. The configuration \textcircled{4} during the flip could be a superposition
(i) \begin{math}
  \simeq \mlr{\qty(
    \Uspin + \Dspin
  )}{2}
\end{math}, of swapping spins, or
(ii) \begin{math}
  \simeq \mlr{\qty(
    \Lspin + \Rspin
  )}{2}
\end{math}, of rotating spins, but is not much of our concern as mentioned in the m.f.\ discussions in sec.\!~\ref{sec:meanfieltheo}.

The next half `\textcircled{5}$\rightarrow$\textcircled{1}' of the loop in fig.\!~\ref{fig:hysteresis} shall be symmetric with the first half. Note the number of \textsc{Onsager}'s order transitions in a loop is four, twice the one of ferroic reversals, which is reminiscent of SU$_2$'s \textit{double covering} of SO$_3$ in a simple group theory.

\subsubsection{An effective single-domain ferroic
\\\textit{Duality to Stoner–Wohlfarth's model in a static case}}
The hysteresis-loop squareness and the rod-bundle morphology are immediate reminiscences of S-W's model for single-domain magnets \cite{tannous2008stoner,carrey2011simple}, with S-W's
(i) macro-spin $\vv{\sigma}$ identified here as projection (at the first configuration) of the sum $\vbm_{\text{sat}}$ of dipolar rods on the anisotropy (indicated by a unit vector $\vba$) axis
(ii) `anisotropy-\textsc{Zeeman}' energy ($\mcE$$_{\text{ani, \textsc{Zee}}} \propto \qty[
  \vba \wedge \mlr{\vv{\sigma}}{\qty|\vv{\sigma}|}
]^2$, $
  - \vv{\sigma} . \vbb
$) competition a mutant of the `nematic-isotropic' entropy competition. Note the above transition properties are implicitly quasistatic, which is relevant at \textit{zero temperature} or \textit{infinite frequency} of $\vbb$, \textit{viz.}, increasing/decreasing the temperature/frequency may \textit{soften} the \textit{hardness} of a ferroic material of strong anisotropy (\textit{cf.}\ the numerical results in fig.\!~3 of \cite{carrey2011simple})

\subsection{Recapture of boundary conditions \& some final remarks}
The ideal case is the cuboid immersed in a r\'eservoir of $\vbb$ absorption, with uniform $\vbb \equiv \vbb^z$ inside and $\equiv \vec{0}$ outside the cuboid. To explain \textit{any} real \cite{speliotis1965hard} and numerical \cite{kuriabova2010linear,jaglajagla2005hysteresis} experiments on strongly anisotropic materials, we shall consider some more realistic conditions. \textit{E.g.},
\begin{enumerate}[label=(\roman*),wide=0pt]
\item rather than the effectively infinite rods through the entire thickness of my model, \cite{kuriabova2010linear} simulated much shorter \textsc{Onsager}'s rods, and defined some power-law two-body potential (eq~(1) of \cite{kuriabova2010linear}) to \begin{enumerate*}
\item soften the rods' hardness or
\item provide bondings between neighbouring ends from different rods, leading to a high-density \textit{columnar} crystal state (an enhanced nematic state; \textit{cf.}\ fig.\!~2 \textit{etc.}\ of \cite{kuriabova2010linear})
\end{enumerate*}
\item \cite{jaglajagla2005hysteresis} models the $\vbm$ of \textit{thin films with easy perpendicular ($z$) axis anisotropy upon change of $\vbb$ along $z$}, with the widths of domain walls much smaller than those of the uniform domains. The system may jump abruptly between $\vbm$ of opposite signs in the hysteresis loops for weak enough exchange/dipolar interactions and Gaussian disorders (introduced to the anisotropy; \textit{cf.}\ figs.\!~2.(9) and 5 \textit{etc.}\ of \cite{jaglajagla2005hysteresis}).
\end{enumerate}
The toy model can thus be seen as a sample (with a finite `height $L$-width $D$' ratio, and $\vbb$'s effective applied area $\sim D^2$) cropped from a uniform domain (say of which the zero-force configuration \begin{math}
  \sim \uspin \!\!\cdots\!\! \uspin
\end{math}) of such a film (whose $\mlr{L}{D_{\text{fil}}} \rightarrow 0$), where the (more or less fixed) boundary conditions at \textcircled{1} and \textcircled{5} might yield some asymmetry `\begin{math}
  \uspin \Uspin \uspin
\end{math} and \begin{math}
  \uspin \Dspin \uspin
\end{math}', but a constant \textit{kink} \cite{chaikin1995principles} energy difference shall not affect significantly the transition physics \textit{etc.}

A last word to mention is that the nonperiodic ferroic model here seems different from the conventional ones of lengthless \textsc{Ising}'s spins (separated by a few crystalline spacings). \textit{Viz.}, some multi-ferroics may be yielded if one could mathematically impose some abstract `infinitely long' ferroic verticals, between which the horizontal impenetrabilities are well protected, compatibly on other periodic ferroic models.

\begin{acknowledgments}
I want to thank (i) the \textsc{sLab} in \textsc{Dongguan} for the hospitality during my gap year in search of a \textit{Ph.D.}
(ii) my lovely classmates in \textsc{Canton} and \textsc{Paris} among others, for occasionally group-studying with me online and for their encouragement to a very scattered soul.
\end{acknowledgments}

\appendix
\section{
\label{app:BLtheo}
Bohr-van Leeuwen's theorem}
The following concepts and derivations are rather general and routine;\footnote{
Natural units such as vacuum lightspeed, \textsc{Planck}'s reduced constant and \textsc{Boltzmann}'s constant are implicitly set to unity. Readers shall also mind the \textit{thermodynamic limit} $C\rightarrow \infty$, and some occasional abuse of notations.
}
one may refer to, \textit{e.g.}, \cite{aharoni2000introduction,pathria2021statistical} or \textsc{Tong}'s online lectures on \href{http://www.damtp.cam.ac.uk/user/tong/dynamics.html}{\textit{classical dynamics}} (2005) and \href{http://www.damtp.cam.ac.uk/user/tong/statphys.html}{\textit{statistical physics}} (2012).
\begin{description}[wide=0pt]
\item[Classical dynamics]{
  specifies a gas of $C$ identical particles (of mass) $\mu$ in a three-dimensional volume $V$ (at a given $\beta \coloneqq 1/$temperature $T$) by a Lagrangian (as a function on a parameterised tangent bundle of a differentiable manifold),
\begin{math}
  \mcL: \mcT_{\mcM} \times \bbR \rightarrow \bbR,
\end{math}
or its \textsc{Legendre} counterpart---a Hamiltonian/\textit{energy}
\begin{align}
  & \mcE\qty[
    \qty{\vbx^i_t}_i, \qty{\vbp^i_t}_i; t
  ] = \sum_i \qty(
    \vbp^i_t = \partial_{
      \partial_t{\vbx}^i_t
    } \mcL
  ) \partial_t{\vbx}^i_t \notag\\ &- \mcL\qty[
    \qty{\partial_t{\vbx}^i_t}_i, \qty{
      \vbx^i_t
      = \qty(x^{i,\alpha}_t)_{\alpha = 1}^3
      \in \mcM
    }_{i = 1}^C; t \in \bbR
  ],
\label{def:hami}\end{align}
with $\vbx^i$, $\vbp^i$ and time derivatives $\partial_t{\vbx}^i$ respectively the (generalised) coordinates, momenta and velocities of $\mu^i$. If $\mu^i$ are charge-$e$ electrons, then an applied magnetic field $\vbB =$ (\begin{math}
  \partial_{\vbx} \wedge
\end{math} vector potential $\vbA$) induces some current densities \begin{math}
  \vbj^i = e \partial_t{\vbx}^i
\end{math}, and some magnetic moments \begin{math}
  \vbm^i = \qty(
    \mlr{\vbx^i \wedge \vbj^i}{2}
) \end{math} whose sum of $\alpha$-directional magnitudes is
\begin{align}
  &M^\alpha \propto \sum_i
  a^{i, \alpha} \partial_t x^{i,\alpha},\text{ with}\!\!\!\!\!\ob{
    \red{\partial_t{\vbx}^i}
    = \partial_{\vbp^i} \mcE
  }^{\text{
    \textsc{Hamilton}'s equation
  }} \!\!\!\!\!\!\xle{\text{`$
    0 \equiv \partial_x q
  $'}} \notag\\ &\ub{
  \blue{\frac{1}{2\mu} \sum_{j}} \partial_{\vbp^i} \blue{\Bigl(}
    \ob{
      \blue{\mu \partial_t{\vbx}^{j}}
      = \vbp^{j} - e\vbA^{j}
    }^{\text{
      minimal coupling
    }} \eqqcolon \wt{\vbp}^{j}
  \blue{\Bigr)^2}
  }_{
    \partial_{\vbp^i} \blue{\qty(
      \mcE - \text{potential energy }
      e U \qty[\qty{\vbx^i}_i]
    )}
  } \red{= \frac{\wt{\vbp}^i}{\mu}},
\label{ham:EMscal}\end{align}
and \begin{math}
  a^{i, \alpha}_q
\end{math} some coefficients of the linear combination.
}
\item[Statistical physics]{
measures the ensemble\footnote{
A virtual collection of many replicas of the statistical system.
} mean
\begin{equation}
  \expval{\mcO}_{
    \mcZ_\beta
  } \xle[
    \dd{\omega} \coloneqq \prod_{i,\alpha} \dd{x^{i,\alpha}} \dd{p^{i,\alpha}}
  ]{
    \bbP^\beta_{\omega} \coloneqq \mlr{
      \mathbb{p}_\omega^\beta
    }{\qty(
      \mcZ_\beta \coloneqq \int
      \dd{\omega'} \mathbb{p}_{\omega'}^\beta
    )}
  } \int \dd{\omega} \bbP_\omega^\beta \mcO_\omega
\label{eq:<O>Z}\end{equation}
of an observable $\mcO$ with \textsc{Boltzmann}'s weights \begin{math}
  \mathbb{p}^\beta_\omega \propto \me^{-\beta \mcE_\omega}
\end{math}, and the normaliser $\mcZ_\beta$ called partition function a \textit{folded rack for displaying statistics}. If $\mcE$ is \blue{the one} in eq~\eqref{ham:EMscal} with no explicit \textsc{Zeeman}'s perturbations `$- \vbm . \vbB$',\footnote{
Quantum effects, as classical electrons cannot whirl forever without collapsing into the nuclei. \textit{I.e.}, quantum mechanics gives $\vbm$ as prior to $\vbB$.
}
then $\mcZ_\beta$ depends trivially on $\vbB$, and any $\vbB$-directional $M^\vbB$ (not necessarily $M^\alpha$ along the basis directions) vanishes averagely.\footnote{
\textit{Viz.}, \begin{math}
  \expval{M^\vbB}
\end{math} is a sum of terms, each of which \begin{math}
  \sim \int \dd{p} \exp{-\mlr{\beta \wt{p}^2}{2 \mu}} \wt{p} \sim 0
\end{math}, as the integrand is odd in $\wt{p}$; \textit{cf.}\ \cite{aharoni2000introduction}. Adding a \textsc{Zeeman} term to $\mcE$, one might expect \begin{math}
  \expval{m} \propto \expval{p} \sim \int \dd{p} \exp{-\qty(\mlr{p^2}{2} - pB)} p \sim \exp{\mlr{B^2}{2}} B \not \equiv 0
\end{math}.
}
}
\end{description}

\section{
\label{app:statprefmaxientr&minifreeener}  
Statistical preference for maximum entropy \& minimum free energy}
\textsc{Shannon}'s \textit{entropy} (a vogue synonym for \textit{disorder})
\begin{equation}
  S \coloneqq \expval{
    -\ln\qty{\bbP_\omega^\beta}
  }_{\mcZ_\beta} \xle{\eqref{eq:<O>Z}}
  - \beta \mcF_\beta \expval{1}_{\mcZ_\beta} + \beta \expval{\mcE}_{\mcZ_\beta},
\label{def:S}\end{equation}
with \textsc{Helmholtz}'s \textit{free energy} (multiplied by $\beta$)
\begin{math}
  \beta \mcF_\beta \coloneqq \qty(- \ln{\mcZ_\beta})
\end{math}
a mathematical equivalent to $\mcZ_\beta$,\footnote{
Summations sometimes present data clearer than multiplications.
} and \begin{math}
  0 \equiv \expval{1}_{\mcZ_\beta} - 1 \eqqcolon \phi_1 
\end{math} a natural constraint of probability. Since \begin{equation}\begin{cases}
  \partial_{\bbP_\omega^\beta} S \equiv \partial_{\bbP_\omega^\beta} \qty( 
    S - \sum_{n=1}^2 \lambda_n \phi_n
    = \lambda_1 . 1 + \lambda_2 \oln{\mcE}
  ) = 0,\\
  \partial_{\bbP_\omega^\beta,\bbP_{\omega'}^\beta} S
  = - \mlr{\qty(
    \delta_{\omega,\omega'}
    = \partial_{\bbP_\omega^\beta} \bbP_{\omega'}^\beta
  )}{
    \bbP_{\omega}^\beta
  } \leq 0,
\end{cases}\end{equation}
$S$ shall maximise (or equivalently, $\beta \mcF_\beta$ tends to minimise) itself for an imposed expectation $\oln{\mcE}$ set by the constraint \begin{math}
  \phi_2 \coloneqq\expval{\mcE}_{\mcZ_\beta}
  - \oln{\mcE} \equiv 0
\end{math}, with the so-called \textsc{Lagrange}'s multipliers \begin{math}
  \lambda_{n=1,2}
  \equiv - \beta \mcF_\beta, \beta
\end{math}. As we only care for energy differences in physics, the sign of $\beta \mcF_\beta$ in eq~\eqref{def:S} may serve as an indicator of some temperature-driven order-disorder transitions. A typical example is \textsc{Peierls}' argument \cite{peierls1936ising} that the paramagnetic disorder overwhelms the ferromagnetic order in \textsc{Ising}'s model if $T$ exceeds a critical temperature $T_\mc$ (\textit{viz.}, thermal fluctuations go wild).\footnote{
\textit{Cf.}\ the simplified description on pp.~44 of \textsc{Bernard} \& \textsc{Jacobsen}'s online lecture manuscript \href{http://www.phys.ens.fr/~dbernard/Documents/Teaching/Lectures_Stat_Field_Theory_vnew.pdf}{\textit{Statistical field theory and applications: an introduction for (and by) amateurs}} (29 June 2020).
} Subject to some tunable forces (\textit{e.g.}, a magnetic field $\vbB$) other than $T$, \textsc{Ising}'s magnets may see some athermal/$T$-fixed transitions.

\section{Miscellaneous of Onsager's model\label{app:misccalc}}
This section is based on \cite[ \textit{etc.}]{parisi2020theory,doi1988theory} and a tutorial of \href{http://lptms.u-psud.fr/membres/trizac/Ens/iCFP/TD2.pdf}{\textit{nematic liquid crystals}} (given by profs. \textsc{van Wijland}, \textsc{Lenz} \& \textsc{Trizac}) I took at \textsc{l'ENS de Paris} in 2019. Consider \begin{equation}
  \mcE = \sum_{i=1}^C \frac{\qty|\vec{p}^i|^2}{2\mu} + \ub{\textstyle \sum_{(i,j)} \qty(
    u_{i,j} \equiv u_{x^{i,j} \equiv \qty|\vec{x}^i - \vec{x}^j|}
    = u_{j,i}
  )}_{\text{sum of interactions by pairs}}
\label{haml:inte_gas}\end{equation}
as a simplest form of eq~\eqref{def:hami} for interacting $\mu^i \equiv \mu$ at a finite concentration $c = C/V$. From eq~\eqref{eq:<O>Z},
\begin{subequations}\begin{eqnarray}
  &&\ub{\textstyle \int \qty(
    \prod_i \frac{\dd{\vec{x}^i}}{V}
  ) \qty(
    1 + \sum_{(j,k)} \msM_{j,k} + \sum_{(j,k),\atop (l,m)} \msM_{j,k} \msM_{l,m} + \cdots
  )}_{
    \text{\textit{cluster expansion} of }\mlr{\qty(
      \mcZ_c \equiv \exp\qty{- \beta f_c V}
    )}{\qty[ 
      \mcZ_c^{\text{ide}} = \mlr{\qty(\mlr{V}{\lambda^3})^C}{C!}
    ]}
  }\qquad\notag\\&&\approx\biggl[ 1 + \ub{\textstyle \qty(
    \int \frac{\dd{\vec{x}}}{2} \msM_x
    \xlongequal{\int \dd{\vec{x}^j} = V} \int \frac{\dd{\vec{x}^i} \dd{\vec{x}^j}}{2V} \msM_{i,j}
  )}_{
    - \text{`excluded volume' }\mcB
  } c \biggr ]^C \\&\Rightarrow\!\!&\beta f_c
  - \ub{\textstyle \qty(
    \beta f^{\text{ide}}_c = c \ln{\frac{c}{c_0}}
  )}_{\text{
    by \textsc{Stirling}'s formula
  }} \approx \!\!\!\!\!\!\!\ub{\textstyle
  c \ln{
    \frac{1}{1 - \mcB c}
  } \approx \mcB c^2
  }_{\substack{
    \text{\textit{two-body scattering} correction}\\
    \text{to the \textit{ideal gas} case `$0 \equiv u$'}
  }}\!\!\!\!\!\!\!,
\label{eq:z/z_idea}\end{eqnarray}\end{subequations}
where (i) \textsc{Mayer}'s function \begin{math}
  \msM_{i,j} \equiv \msM_{x^{i,j}}
  \coloneqq \qty(\me^{-\beta u_{i,j}} - 1)
\end{math} is assumed to be mostly small\footnote{
\textbf{NB} \cite{parisi2020theory}\;
(i) the so-called \textsc{Virial}'s expansion~\eqref{eq:viriexpa-Pi} or \eqref{eq:z/z_idea}, \eqref{freenrgb:onsa} is formally only reliable at $\beta$, $c \rightarrow 0$
(ii) for perfectly hard particles with \begin{math}
  u_{x^{i,j} \leq l} \equiv \infty
\end{math} and \begin{math}
  u_{x^{i,j} > l} = 0
\end{math} (say $\vbx^j$ sees the excluded volume of $\vbx^i$ when \begin{math}
  x^{i,j} \leq l
\end{math}), \begin{math}
  \msM_{i,j}
\end{math} is really some \textsc{Heaviside}'s step function \begin{math}\qty(
  - \vartheta_{l-x^{i,j}}
)\end{math} and independent of $\beta$.
\label{foot:Mayefct}}
(ii) $c_0$ and \textsc{de Broglie}'s wavelength \begin{math}
  \lambda = \sqrt{\mlr{2\pi \beta}{\mu}}
\end{math} are irrelevant reference constants
(iii) we would like to express everything as a function of $c$ up to $\mcO\qty(c^2)$. \textit{E.g.},
\begin{eqnarray}
  &&\!\!\ub{\textstyle
    \beta \Pi \coloneqq - \partial_V \eval{\!\!\big(
    \beta f_c V
    \textstyle}_C
  }_{\textit{osmotic pressure}} = \ub{\textstyle\qty[
    \beta u^{\text{che}} \coloneqq \partial_c \eval{\!\!\big(
      \beta f_c
    \textstyle}_V
  ]}_{\textit{chemical potential}} \ub{
    \eval{\!\!\big(\!
      -\!V \partial_V c  
    \textstyle}_C
  }_{
    c
  } - \beta f_c \quad\notag\\&&\!\!\approx
  c + \qty[ 
    2 \mcB - \mlr{\qty(\beta f_c + c)}{c^2}
    \approx \mcB
  ] c^2.
\label{eq:viriexpa-Pi}\end{eqnarray}
If $\mu^i$ are now \textsc{Onsager}'s rods pointing in directions (of some unit vectors) $\vbn^i$, one might promote the scalar $c$ into a vector $\vbc = \qty(c \Psi_\vbn)$ by adding a factor of orientational probability distribution $\Psi_\vbn$ constrained by
\begin{subequations}\label{freenrg:onsa}\begin{equation}
  1 = \int \dd{\vbn} \Psi_{\vbn},
\label{freenrga:onsa}\end{equation}
and reformulate eq~\eqref{eq:z/z_idea} directly into\footnote{
\textsc{Lagrange}'s multiplier $\lambda$ for the normalisation~\eqref{freenrga:onsa} is redundant once an explicit form~\eqref{dist:Psi(alph)} of \begin{math}
  \Psi_\vbn
\end{math} is given.
}
\begin{align}
  &\beta f_\vbc =\ub{\textstyle
  \beta f_c^{\text{ide}}
  + c \int \dd{\vbn} \Psi_{\vbn} \ln{\Psi_{\vbn}}
}_{
  \int \dd{\vbn} \vbc\ln\qty{\mlr{
    \vbc
  }{c_0}}
} + \ub{\textstyle
  c^2 \int \dd{\vbn^i} \dd{\vbn^j} \Psi_{\vbn^i} \Psi_{\vbn^j}
}_{
  \int \dd{\vbn^i} \dd{\vbn^j} \vbc^i \vbc^j
}
\notag\\&\times \qty(\mlr{
  \mcB_{\vbn^i,\vbn^j} \equiv - \int \dd{\vec{x}^{i,j}} 
  \msM_{\vbn^i,\vbn^j;x^{i,j}
}}{2}),
\label{freenrgb:onsa}\end{align}\begin{equation}
  0 = \ub{\textstyle
    \lambda + \beta \qty(
      1 + \ln{\Psi_{\vbn^i}}
      + \int \dd{\vbn^j} \Psi_{\vbn^j} 2 \mcB_{\vbn^i,\vbn^j} c
  ) c}_{\fdv*{\qty[ 
    \lambda \eqref{freenrga:onsa} + \eqref{freenrgb:onsa}
  ]}{\Psi_{\vbn^i}}}.
\label{freenrgc:onsa}\end{equation}
% \begin{align}
%   &\underbracket{\textstyle
%   0 = \fdv{\qty(
%     \lambda \int \dd{\vbn} \Psi_{\vbn} + f\qty[\vec{c}]
%   )}{\Psi_{\vbn_i}}
%   }_{\text{MF condition}} = \lambda +
%   \notag\\ & c \qty(
%     1 + \ln\qty{\Psi_{\vbn_i}}
%     + 2 \int \dd{\vbn_j} \Psi_{\vbn_j} c \mcB \qty(\vbn_i,\vbn_j)
%   ).
% \label{freenrgc:onsa}
% \\&\xRightarrow[\eqref{freenrga:onsa}]{\eqref{freenrgb:onsa}} \underset{
%   \Psi
% }{\text{min}}\qty{\beta f\qty[\vec{c}]} \xlongequal{\lambda = 0}
% \notag\\ & c \biggl(
%   \ln\qty{\frac{c}{c_0}} - 1
% \underbracket{\textstyle
%   - \int \dd{\vbn_i} \dd{\vbn_j} \Psi_{\vbn_i} \Psi_{\vbn_j} c \mcB \qty(\vbn_i,\vbn_j)
% }_{\mlr{
%     \int \dd{\vbn} \Psi_{\vbn} \qty(\ln\qty{\Psi_{\vbn}}
%     + 1
%   )
% }{2}}\biggr).\notag
% \end{align}
\end{subequations}
\textsc{Onsager} proposed a trial distribution
\begin{equation}
  \Psi^{
    \text{order parameter }s
  }_{\theta = \cos^{-1}\qty{\vbn . \vbn^\text{reference}}} \propto \frac{\cosh\qty{s \cos{
    \theta
  }}}{\mlr{4\pi \sinh\qty{s}}{s}},
\label{dist:Psi(alph)}\end{equation}
with $s$ chose (by the system) to minimise eq~\eqref{freenrgb:onsa} (\textit{i.e.}, to reach \begin{math}
  \beta f_{c, s = s^{\text{sad}}}
  % \equiv \underset{s}{\text{min}}\qty{\beta f}
\end{math} by seeking solutions to the m.f.\ condition/saddle-point eq~\eqref{freenrgc:onsa}) and to parameterise the order: \textit{isotropic}\begin{math}
  \xleftrightarrow{\text{phase}}
\end{math}\textit{nematic} $\Leftrightarrow$ \textit{flat~$\forall \theta$}\begin{math}
  \xleftrightarrow[\Psi_{\theta} = \Psi_{-\theta}]{\text{shape of }\Psi}
\end{math}\textit{peaks as \textsc{Dirac}'s~$\delta$ at $\theta = \mathbb{Z} \pi$ and vanishes elsewhere }$\Leftrightarrow$ $0$\begin{math}
  \xleftrightarrow[
    \Psi^{s} = \Psi^{-s}
  ]{
    \qty|s|
  }
\end{math}$\infty$. Substituting further some hypothetical \begin{math}
  \mlr{2\mcB_{\vbn^i,\vbn^j} c}{\varphi}
  \sim \qty|\vbn^i \wedge \vbn^j|
\end{math} into eq~\eqref{freenrgb:onsa}, one could read off useful m.f.\ information about the preferable \textit{single} phase from the diagrams of the energy differences \begin{math}
  \delta f_{\varphi} [s] \coloneqq \beta f_{\varphi} \qty[s \equiv s_{\text{nem}}] - \beta f_{\varphi}\qty[0 \equiv s_{\text{iso}}]
\end{math}.\footnote{
Which is enough for the scope of this essay. To detail the phase coexistence \textit{etc.}, one shall work out \begin{math}
  f_{
    s^{\text{sad}}_{\text{iso}}
  } \qty[\varphi]
\end{math} and \begin{math}
  f_{
    s^{\text{sad}}_{\text{nem}}
  } \qty[\varphi]
\end{math} from eq~\eqref{freenrg:onsa}, and apply a common tangent to the two curves (at points \begin{math}
  \varphi^*_{\text{iso, nem}}
\end{math} respectively, with \begin{math}
  \varphi \in \qty[
    \varphi^*_{\text{iso}},\varphi^*_{\text{nem}}
  ]
\end{math} the phase coexistence region; \textit{cf.}\ fig.~10.2 of \cite{doi1988theory}). For some more general knowledge of the crossover behaviours (spinodal decomposition, nucleation \textit{etc.}), \textit{cf.}, \textit{e.g.}, \cite{porter2009phase} (for an introduction to thermodynamics in materials) or \cite{parisi2020theory} (for a pedagogical description of escaping a metastable state through nucleation, where the system of \textsc{Ising}'s magnets with short-range interactions could stay in a metastable state for a lifetime \begin{math}
  \tau \sim \me^{\delta{\qty(\beta \mcF)}}
\end{math}, with $\delta{\qty(\beta \mcF)}$ the surface tension energy of the phase separation region).
}
Since
(i) all we need here is a simple illustration
(ii) the direct analytical/numerical computations for the m.f.\ energies in \textsc{Onsager}'s model could be quite involved \cite{vroege1992phase}, try some makeup series (at the request of a minimal description for discontinuous transitions like `gas condensation to liquid')
\begin{subequations}\begin{equation}
  \delta f_\varphi \qty[\sigma] = g^2_\varphi \sigma^2 - g^3_\varphi \sigma^3 + g^4_\varphi \sigma^4 + \mcO\qty(\sigma^5)
\label{eqa:f[sigma]}\end{equation}
of \textsc{Landau}'s to mimic the one schematically shown in fig.\!~10.1 of \cite{doi1988theory},
with $\sigma_s$ some makeup order parameter\footnote{
A usual convention, \begin{math}
  \sigma_s \coloneqq
  \expval{\msP^2_{\cos{\theta}}}_{\Psi^{s}}
  \xle{\eqref{dist:Psi(alph)}} 1 - \mlr{3}{s\tanh{s}} + \mlr{3}{s^2}
  \xrightarrow{s \rightarrow 0} \mlr{s^2}{15} + \mcO\qty(s^4)
\end{math}, in liquid crystals \cite{chaikin1995principles}, with $\msP^2$ \textsc{Legendre}'s degree-$2$ polynomial.
\label{foot:sigm[alph]}} 
equivalent to $s$. To capture physics in the vicinity of a transition point, assume a linear form \begin{math}
  g^2_\varphi = g^0 - g^1 \varphi
\end{math} (as we expect some sign changes of the $\sigma^2$-term due to the variations of $\varphi$), with \begin{math}
  g^{0,1}
\end{math} as well as the higher-order coefficients $g^3$ (with the prefixed minus sign to induce a local maximum) and $g^4$ constants. Up to some rescaling (to have some desired form \begin{math}
  \qty[ \sigma_{1,2}^*,\varphi_{1,2}^* ] \equiv [\sigma_1^*,0],
  [0,1]
\end{math} \textit{etc.}),
\begin{equation}
  \ub{\textstyle
    \delta \wt{f}_{\wt{\varphi}} \qty[\wt{\sigma}]
  }_{
    \mlr{\delta f_\varphi \qty[\sigma]}{\oln{f}} 
  } = \ub{\textstyle
    \qty[\mlr{\qty(
      1-\wt{\varphi}
    )}{2}]
  }_{
    \mlr{g^2_\varphi \oln{\sigma}^2}{\oln{f}}  
  } \wt{\sigma}^2 - \ub{\textstyle
    \qty(\mlr{2}{3})
  }_{
    \mlr{g^3 \oln{\sigma}^3}{\oln{f}} 
  } \wt{\sigma}^3 + \ub{\textstyle
    \qty(\mlr{1}{4})
  }_{
    \mlr{g^4 \oln{\sigma}^4}{\oln{f}} 
  } \wt{\sigma}^4,
\label{eqb:f[sigma]}\end{equation}\label{eq:f[sigma]}\end{subequations}
with the tildes simply dropped hereafter;
\begin{figure}[htbp]
\includegraphics[width=.33\textwidth]{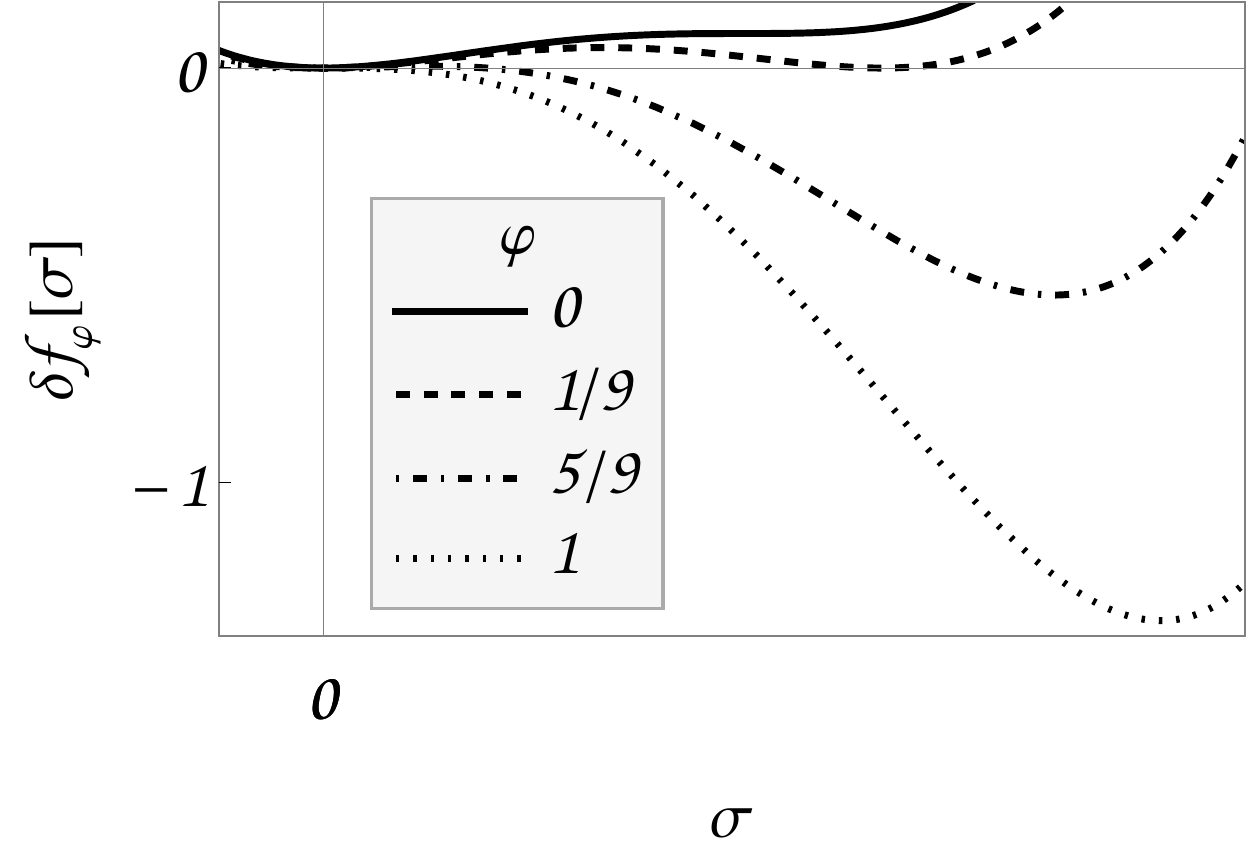}\end{figure}
(i) saddle-point condition \begin{math}
  0 = \partial_\sigma \delta{f}_\varphi
  \xRightarrow{\varphi \in [0,1]}
\end{math} two local-minimum solutions
\begin{equation*}
  \ub{\textstyle
    \delta{f}_\varphi \qty[\sigma = 0] = 0
  }_{\text{fixed}},\;
\ub{\textstyle
  \delta{f}_\varphi \qty[\sigma = 1 + \sqrt{\varphi}]
  = \frac{1}{12} - \qty(
    \frac{1}{2} + \frac{2\sqrt{\varphi}}{3} + \frac{\varphi}{4}
  ) \varphi
  }_{
    \sim \mcO\qty(\varphi^2)
    \text{ indeed}
  },
\end{equation*}
and a local-maximum one \begin{math}
  \sigma = 1 - \sqrt{\varphi}
\end{math}
(ii) \begin{math}
  0 = \partial_\sigma \delta{f}_\varphi
  = \partial^2_{\sigma} \delta{f}_\varphi
  \Rightarrow
\end{math} two reflective points \begin{math}
  \qty(\sigma_{1,2}^*,\varphi_{1,2}^*)
  = (1,0), (0,1)
\end{math}, where the metastable nematic/isotropic state appears respectively when increasing/decreasing $\varphi$
(iii) \begin{math}
  \qty(0 \equiv \delta{f}_\varphi \qty[0])
  = \delta{f}_\varphi \qty[\sigma \neq 0]
  \xRightarrow{0 = \partial_{\sigma} \delta{f}_\varphi}
\end{math} a transition point \begin{math}
  \qty(\sigma_\mc^*,\varphi_\mc^*) = \qty(\mlr{4}{3},\mlr{1}{9})
\end{math}. In short,
(i) the isotropic/nematic state is more stable respectively for \begin{math}
  \varphi \in \qty(\varphi_1^*,\varphi_\mc^*)
\end{math} or \begin{math}
  \in \qty(\varphi_\mc^*,\varphi_2^*)
\end{math}
(ii) \begin{math}
  \delta{f}_\varphi
\end{math} has a unique minimum dubbed isotropic/nematic respectively for \begin{math}
  \varphi < \varphi_1^*
\end{math} or \begin{math}
  > \varphi_2^*
\end{math}.

\end{document}